\documentclass[12pt]{article}

\pdfoutput=1

\usepackage{graphicx,amsmath}
\usepackage{units}

\usepackage{xcolor}
\usepackage[utf8x]{inputenc}
\usepackage{float}
\newlength{\dinwidth}
\newlength{\dinmargin}
\def\lapproxeq{\lower .7ex\hbox{$\;\stackrel{\textstyle                                                    
<}{\sim}\;$}}                                                    
\def\gapproxeq{\lower .7ex\hbox{$\;\stackrel{\textstyle                                                    
>}{\sim}\;$}}                                                    
\def\be{\begin{equation}}                                                    
\def\ee{\end{equation}}                                                    
\def\bea{\begin{eqnarray}}                                                    
\def\eea{\end{eqnarray}}

\def\sh{\hat s}
\def\sh2{{\hat s}^2}
\def\PDF{{\rm PDF}}

\newcommand{\dd}{\mathrm{d}\, } 
\begin{document}
%\titlepage

\begin{flushright}                                                    
                                                  
\today \\                                                    
\end{flushright} 

\vspace*{0.5cm}

\begin{center}
{\Large \bf Improving the Drell--Yan probe of small $x$ partons \\
at the LHC via an azimuthal angle cut}

\vspace*{1cm}
                                                   
M.~Kendi Yamasaki$^{a}$, Emmanuel G.~de Oliveira$^{a}$  \\                                                    
                                                   
\vspace*{0.5cm}                                                    
$^a$ {\it Departamento de F\'{i}sica, CFM, Universidade Federal de Santa
	Catarina, C.P. 476, CEP 88.040-900, Florian\'opolis, SC, Brazil}\\        
                                              
\vspace*{1cm}  

%-----------------------------------------------------------------
                                                    
\begin{abstract}                                                   

Predictions for Drell--Yan lepton pair production at low dilepton mass and small $x$ at the LHC usually have a large scale dependence. This can be decreased by determining an \textit{optimal factorization scale}. In this paper, we reduce this scale by imposing a cutoff in azimuthal angle between the transverse momentum of the leptons, properly taking into account Sudakov effects. This allows one to probe the parton distributions at smaller scales eliminating most of the current theoretical  uncertainty.
\end{abstract}                                                          
\vspace*{0.5cm}                                                    
                                                    
\end{center}                                                    

%-----------------------------------------------------------------
\section{Introduction}

The Drell--Yan process is one of the standard channels for determining the parton distribution functions (PDFs), specially the sea quark ones. At the CMS experiment, for instance, the production of the pairs of muons is measured with a wide range of dilepton invariant mass, $15 < M < 3000$ GeV at $\sqrt{s} = 13$ TeV~\cite{Sirunyan:2018owv}. The results are integrated in dilepton rapidity and show good agreement with next-to-next-to-leading order of Drell--Yan predictions. For a similar result from ATLAS, see Ref.~\cite{Aad:2019wmn}. 
 
It is possible to calculate the Drell--Yan (DY) cross section through a factorized scheme: it is  as a convolution of the parton distributions (one for each involved proton) with the matrix element using a factorization scale, $\mu_F$. Schematically, we have: 
\be \label{eq:had_cross_sec} \sigma = \int\dd x_A \dd x_B ~\PDF(x_A, \mu_F) \times |\mathcal{M}(\mu_{\rm F})|^2\times  \PDF(x_B, \mu_F)~,\ee
where the matrix element, $\mathcal{M}(\mu_{\rm F})$, is calculated in a perturbative manner. The convolution is in $x$ space, i.e., the longitudinal momentum fraction carried by the partons. At leading order (LO), there is a big scale dependence, whereas, at next-to-leading order (NLO), there is a smaller dependence and so on for higher orders until that, if we consider all perturbative terms, the result would be independent of scale, assuming it is a convergent series. The conventional choice for the factorization scale is $M$ for the DY process. 

\begin{figure}[tb]
	\begin{center}
		\includegraphics[width = 5 cm]{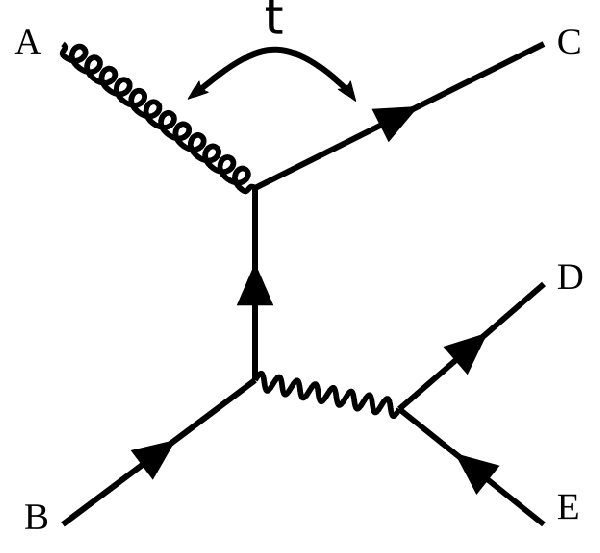}
		\caption{\label{fig:diagram} The Compton scattering diagram of the NLO Drell--Yan process: gluon A and quark B are the initial particles, resulting in a quark C and a photon, which in turn splits into a pair of leptons D and E. This diagram has a divergence in the $t$ channel and it is the most relevant one  at NLO for small $x$ due to the gluon distribution.}
	\end{center}
\end{figure}

It is known that at small $x$ NLO theoretical predictions there is a large factorization scale dependence, usually quantified by allowing for $\mu_F = M/2, 2M$. This is due to the fact that a variation of factorization scale will change the parton distributions. If the whole perturbative series were present, the matrix element would cancel this change. However, when truncated at NLO, the matrix element contains only one parton emission (see, e.g., Fig.~\ref{fig:diagram}), while the parton distributions can emit many partons (average of 8 at small $x$ and for the LHC energies, as estimated in Ref.~\cite{deOliveira:2012ji}) when they are evolved in $\mu_F$. This uncertainty limits the precision in which the parton distributions can be probed by the Drell--Yan process.  

However, there is a procedure~\cite{deOliveira:2012ji} to set an \textit{optimal scale}, which reduces the uncertainty due to the factorization scale. The main idea is, in the limit of small $x$,  to include part of the NLO contribution already at the LO by changing the parton distribution factorization scale at LO. It was applied first for the DY process, but has been also applied to other processes like: $c\bar{c}$ and $b\bar{b}$ production \cite{deOliveira:2016bvs} and $J/\psi$ production \cite{Flett:2019pux}. 

Given that at large scales the parton distributions are more or less understood, it would be desirable to lower the optimal scale. With this goal, in Ref.~\cite{deOliveira:2012mj}, a dilepton (or, equivalently, photon) upper transverse momentum ($k_t$) cutoff was imposed, therefore making the NLO contribution smaller and then requiring a smaller optimal scale. In this way, one has information about the PDFs at smaller scales, i.e., smaller than the scales that can currently be measured due to experimental limitations. In this paper, we continue that work by imposing a cut in the azimuthal angle between the transverse momentum of the leptons (instead of a photon $k_t$ cut). This will be a complementary approach, that can be tested both theoretically and experimentally, if measured. 

There are other ways to ressum small $x$ parton evolution. For example, an all-order small-$x$ resummation matched to a fixed order DGLAP anomalous dimension \cite{Marzani:2015oyb} was obtained some time ago. Also, by considering the perturbative coefficient functions at fixed-order minus its expansion in $\alpha_s$ series, it was possible to resum small $x$ effects in Refs.~\cite{Bonvini:2017ogt, Abdolmaleki:2018jln} and have a better description of DIS data. In our work, we have the advantage of being able to choose more exclusive observables by having an easier way of introducing cutoffs.

\begin{figure}[bt]
	\begin{center}
		\includegraphics[width = 6 cm]{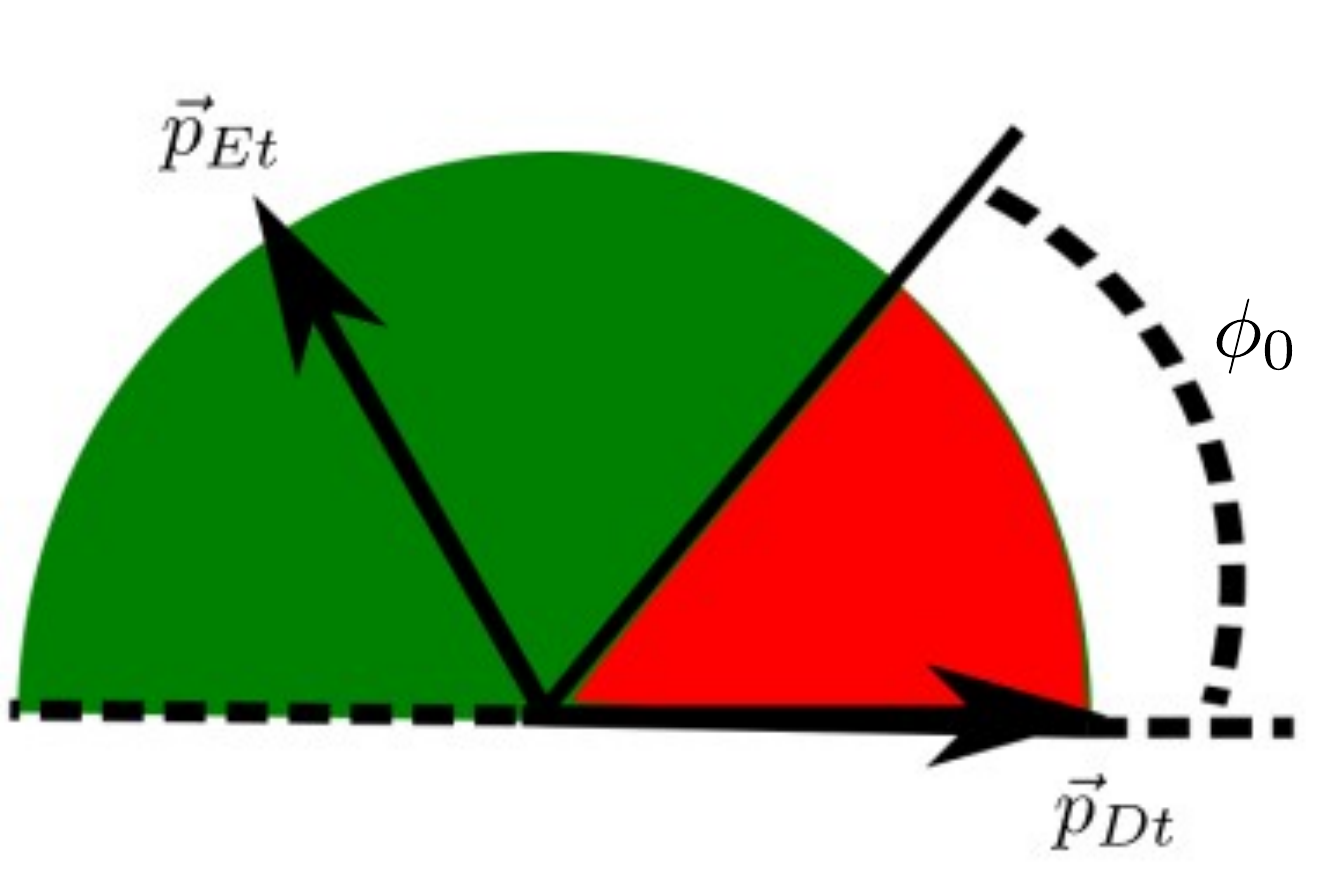}
		\caption{\label{fig:angle}Transverse momentum vector of leptons D and E, separated by an azimuthal angle $\phi$. For a given cutoff $\phi_0$, the green region corresponds to allowed values of angles $\phi > \phi_0$, i.e. the part of phase space which is taken into account in the calculation. The red region is cut off, therefore, the events that are closer to the back-to-back configuration are the relevant (measured) ones.}
	\end{center}
\end{figure} 

This paper is organized as follows: in the Sec.~\ref{sec:angle cut}, we discuss how to reduce the NLO phase space through the azimuthal angle cut. Then, in Sec.~\ref{sec:imposig cut}, we calculate the optimal scale as a function of the cutoff. In Sec.~\ref{sec: effect of cut in the cross section}, we show the effect of the cutoff in the cross section and, in Sec.~\ref{sec: sensitivity}, we show the stability of the results with regard to the choice of the remaining factorization scale. Finally we present our conclusions in Sec.~\ref{Conclusion}.

%-----------------------------------------------------------------
\section{Imposing an azimuthal angle cut $\phi_0$}
\label{sec:angle cut}

Drell--Yan process at NLO is given by a collision between a parton A and parton B, resulting in another parton C and a photon, the latter splitting into leptons D and E. The most important case at small $x$, where the gluon distribution dominates, is the QCD Compton scattering: a gluon and a quark are the initial partons that result in the quark C and the leptons D and E, as shown in Fig.~\ref{fig:diagram}. We define the Mandelstam variable $t = (p_C - p_A)^2$.

The leptons D and E with the corresponding transverse momentum $\vec{p}_{Dt}$ and $\vec{p}_{Et}$  are separated by an azimuthal angle $\phi$. If we take $\phi$ to be the smallest angle, it will vary between $0 < \phi < \pi$, with the upper limit corresponding to the back-to-back configuration. With an azimuthal angle cut, we reduce the number of events taken into account by selecting only the ones with $\phi > \phi_0$, i.e., closer to the back-to-back configuration. In Fig.~\ref{fig:angle}, we present only the lepton pair, D and E, and show the cut off region in red.

By introducing the cutoff, we expect to lower optimal scale that will be described in the next section. In this way, we are able to safely probe parton distribution at lower scales by reducing the  big uncertainty involved in the choice of this scale as shown in Fig.1 of the Ref.~\cite{deOliveira:2012mj}.

%-----------------------------------------------------------------
%Sudakov form factor
\section{Determination of the optimal scale}
\label{sec:imposig cut}

Following the procedure of Ref.~\cite{deOliveira:2012mj}, we use the parton cross section for the NLO subprocess $qg\rightarrow q \gamma^* \rightarrow q l \overline{l}$ differential in $M^2$, in $t$ and in the lepton transverse momenta. This is integrated in the two lepton variables, keeping the restriction in the azimuthal angle of $\phi> \phi_0$:
\begin{align}
\int \dd p_{Dt} \int \dd p_{Et} \frac{\dd \hat{\sigma}(qg \rightarrow q l \overline{l})}{\dd M^2 \dd t \dd p_{Dt} \dd p_{Et}} \Theta(\phi -\phi_0)
\end{align}
We also use the LO parton cross section convoluted with DGLAP $g\rightarrow q \overline{q}$ splitting function~\cite{DGLAP} that does not have a dependence on the lepton variables:
\begin{align}
\frac{\alpha^2 \alpha_s z}{9 M^4} \frac{z^2 + (1-z)^2}{t}.
\end{align}

We equate both expressions (NLO vs. LO convoluted with DGLAP) and integrate in $t$. The infrared divergences cancel (the cut does not touch the divergence). There is a further integration in $z = M^2/\hat{s}$ with fixed $M$, accounting for an incoming gluon flux of $1/z$, where the parton c.o.m. energy is $\sqrt{\hat{s}}$. Thus, we have an equation that can be used for finding the optimal scale, $\mu_0$.  

In the next step, to calculate the cross section, we will use the factorized scheme:
\begin{equation}\label{eq:sigma_lo_nlo}
    \sigma = \rm PDF (\mu_0) \times C^{LO} + PDF(\mu_F)\times C^{NLO}(\mu_0) 
\end{equation}
using the optimal scale, $\mu_0$, in the parton distribution appearing at leading order and also in the next-to-leading order coefficient, $C^{\rm NLO}$.
%The main effect of the procedure described above is to include most of the NLO contribution already at the leading order by changing the factorization scale of the parton distributions convoluted with the LO matrix element. 
By using the optimal scale $\mu_0$ we include in the LO term all the NLO contributions which depends on factorization scale and enhanced by a large $\ln(1/x)$ -- that is we resum inside the LO low-$x$ PDF the terms $[\alpha_s\ln(\mu_F/M)\ln(1/x)]^n$. Of course now, the first of these terms should not be taken into account at NLO to avoid double counting; this is done by setting $\mu_F = \mu_0$ in $C^\text{NLO}$.

However, since there is a cutoff applied, it is necessary to take care of the situation of a parton that emits other partons during the evolution that may spoil the cutoff. In other words, we must take into account possible parton emissions from the optimal scale ($\mu_0$) up to the hard scale ($\sqrt{\hat{s}}$) which give a supplementary transverse momentum to the dilepton. For example, a configuration in which the leptons are exactly back-to-back ($\phi= \pi$) when the dilepton has no transverse momentum can be changed to another configuration like $\phi = \pi/2$ if the dilepton is given the appropriated transverse momentum. We will do that at double logarithm accuracy. 

This situation is addressed by including Sudakov form factors that assure there will be no emission between the optimal scale $\mu_0$ and $\sqrt{\hat{s}}$. This inclusion is detailed in Ref.~\cite{deOliveira:2012mj}, here we briefly recall that, in the double log approximation, the quark Sudakov factor is given by:
\begin{equation}
    T_q = \exp{(-\alpha_s S_q(\mu_0,\sqrt{\hat{s}}))}
\end{equation}
with
\begin{equation}
    S_q = \frac{C_F}{\pi} \Re \left( \ln(\sqrt{\hat{s}}/\mu_0) +i\pi/2 \right) ^2
\end{equation}
where $C_F = 4/3$ and, at leading order, $\sqrt{\hat{s}} = M$. Similarly, there is a Sudakov factor for the gluon. They enter the Eq.~\ref{eq:sigma_lo_nlo} as factors that multiply respective the parton distributions. Of course now we have to exclude the first term $\alpha_s (\ln^2(\sqrt{\hat{s}}/\mu_0) - \pi^2/4)$ from the $C^\text{NLO}$ expression to avoid the double counting.

One may argue that it is not clear how the Sudakov factors could be used with the angular cut, since they are traditionally used to account for no emission in a range of transverse momentum. First of all, the Sudakov factor depends on virtualities of single particles (as in the original paper \cite{Sudakov:1954sw}), not transverse momentum. This means that we can use them here, provided that we use as their arguments the inclusive scale $M$ or $\hat{s}$, where all possible dileptons are taken into account,  and the optimal scale $\mu_0$ with our cutoff. This is good at double logarithm accuracy and corrections to it will appear only at NNLO. As shown in Ref.~\cite{deOliveira:2012ji}, the NNLO is rather small after the choice of the optimal scale and that justifies our approach. If we were to make completely sure that the cutoff was not spoiled by the PDF evolution, we would have to calculate this process to all orders or do a Monte Carlo evolution keeping track of all variables of intermediate partons, but we do not pursue this complicated approach.

\begin{figure}[tb]
	\begin{center}
		\includegraphics[width = .8\textwidth]{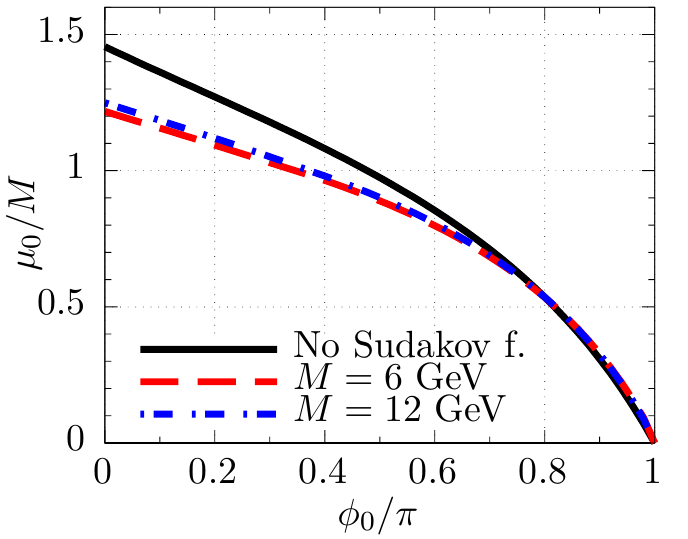}
		\caption{\label{fig:mu0} Optimal scale as a function of the azimuthal angle cutoff with and without Sudakov factors. We observe that for $\phi_0 > 0.7 \pi$ the Sudakov factor does not affect much the optimal factorization scale.}
	\end{center}
\end{figure}

In the Fig.~\ref{fig:mu0}, it is shown the reduction of the optimal scale with the cutoff for the cases without and with  Sudakov form factor, for dilepton masses equal to 6 and 12 GeV. It starts with the case of no cut applied ($\phi_0=0$) and ends in the most drastic case of $\phi_0 = \pi$, where all phase space is cut off. In this range, the optimal scale varies from $\mu_0 = 1.45\,M$ (no cutoff) to $\mu_0 = 0$. 

From Fig.~\ref{fig:mu0}, we clearly see that, in the region which starts around $\phi_0 = 0.7\pi$, Sudakov effects are not so important on the determination of the optimal factorization scale. This is the most important region to study smaller scales, since $\mu_0/M < 0.7$ in this case. Then, we can investigate predictions of Drell--Yan cross section at smaller scales without worrying about a new theoretical uncertainty due to the Sudakov form factors. 

After including into the LO term most of the NLO contribution, it would still be possible to use the parton distributions at a different scale $\mu_1$ when computing the NLO contribution. Then the NNLO coefficient would depend on $\mu_0$ and $\mu_1$ and the idea would be to choose $\mu_1$ in a way to make that almost all of the NNLO contribution would be already taken into account at lower orders. This would further reduce the scale uncertainty. We do not pursue such calculation here, but we argue, as first discussed in Ref.~\cite{deOliveira:2012ji}, that setting $\mu_1 = \mu_0$ already is a good choice, since the dominant diagram at small $x$ at NNLO is the one with two gluons in the initial state and most of its contribution will be taken into account by correcting both quark and antiquark legs of the LO diagram with LO (and not NLO) DGLAP. 

Another possibility is to combine the azimuthal angle cutoff with the transverse momentum cutoff discussed in Ref.~\cite{deOliveira:2012mj}. This would lower the optimal scale w.r.t. the application of a single cut, but we expect it will not be much lower. In fact, we expect that both cutoffs will be similar in the sense that a large part of the phase space is cut by the two cuts. For instance, the optimal scale for $\phi_0 = 0.85 \pi$ is $\mu_0 = 0.44 M$; if we also cut the dilepton transverse momentum at $k_0 = M$, the optimal scale is still $ 0.44 M$ within rounding error, if we set  $k_0 = M/2$, we have $\mu_0 = 0.42 M$. In conclusion, applying both cuts should be weighted against the possible experimental difficulties when measuring this new cross section, depending on the setup it will be better to apply a single but stricter cut.

%-----------------------------------------------------------------
\section{Predictions with an azimuthal angle cutoff}
\label{sec: effect of cut in the cross section}

\begin{figure}[tb]
	\begin{center}
		\includegraphics[width = .8\textwidth]{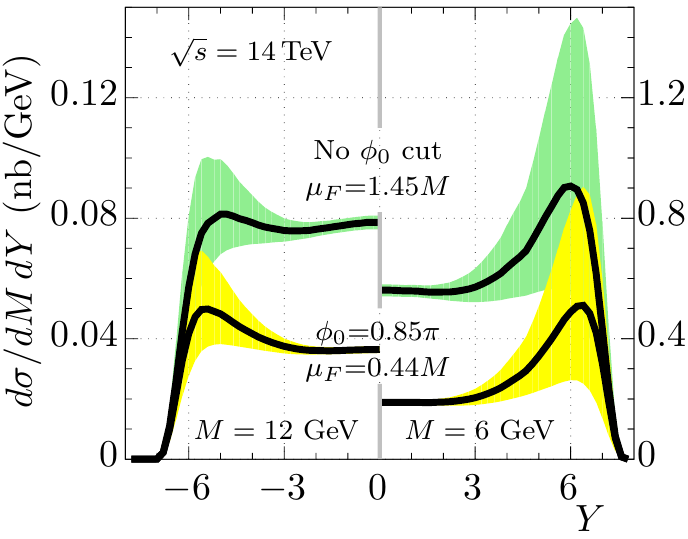}
	\end{center}
	\caption{\label{fig:uncert} Drell--Yan differential cross section for $M = 12$\,GeV (left) and $6$\,GeV (right). The upper curves correspond to the result without any cut ($\mu_0 = 1.45M)$, while the lower ones, to the result with an azimuthal angle cutoff of $\phi_0 = 0.85\pi$. The bands display the $1\sigma$ PDF uncertainty and show that they can be reduced by a proper measurement (with current LHC precision) of such observable.}
\end{figure} 

As described in the Secs.~\ref{sec:angle cut} and \ref{sec:imposig cut}, we are now in position to lower the scale with an azimuthal angle cut and investigate the effects of the cutoff in cross section. We are interested in applying a cut for which Sudakov factors do not change much our results, $\phi_0 > 0.7\pi$. A good choice will be  $\phi_0 = 0.85\pi$, for which the optimal scale is reduced to $\mu_0 = 0.44 M$. In Fig.~\ref{fig:uncert}, we show our predictions for the differential cross section in dilepton rapidity $Y$ for the Drell--Yan process at LHC energy of $\sqrt{s}= 14$ TeV. We use MMHT14 NLO PDFs~\cite{Harland-Lang:2014zoa} and set the dilepton mass equal to 6 and 12\,GeV. 

The upper curves in Fig.~\ref{fig:uncert} correspond to the absence of any cutoff; therefore  $\mu_F = \mu_0 = 1.45 M$. In this case, the scale at which the partons are probed is still larger than the usual choice $\mu_F = M$. The lower curves correspond to the cutoff $\phi_0 = 0.85\pi$, for which we have a much lower scale (less than a third of $1.45M$), but we still have a considerable cross section, as it can be seen that approximately 50\% of the dileptons produced are kept. 

We also calculate the $1\sigma$ error corridors coming from the PDF uncertainty, that, depending on $Y$, are rather large. The current precision of the measurements at the LHC is better than this PDF uncertainty, leading us to believe that a proper measurement of such observable would add new precise knowledge about the PDFs. In the next section we will see that the remaining factorization scale will be smaller than such bands. 

%-----------------------------------------------------------------
\section{Sensitivity of choice of factorization scale}
\label{sec: sensitivity}

\begin{figure}[tb]
	\begin{center}
		\includegraphics[width = .8\textwidth]{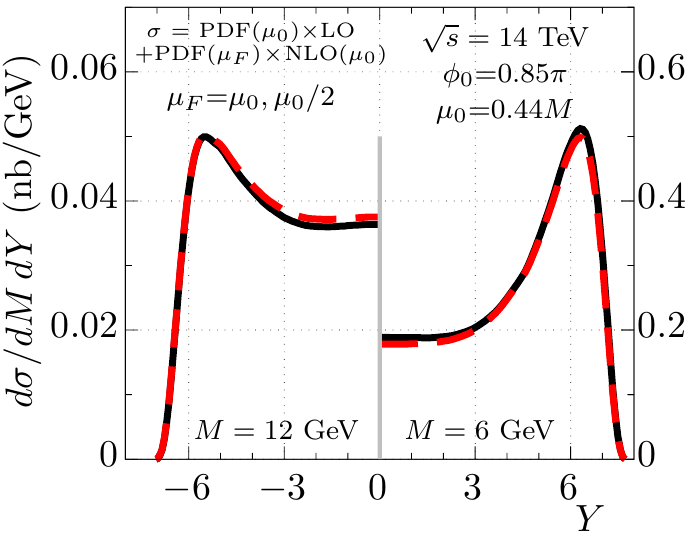}
		\caption{\label{fig:sensitivity} Drell--Yan differential cross section given at two  factorization scales $\mu_F = \mu_0$ (black) and $\mu_0/2$ (red). The azimuthal angle cutoff $\phi_0 = 0.85 \pi$ is imposed, with optimal scale at LO given by $\mu_0 = 0.44 M$. This shows that the remaining factorization scale uncertainty is greatly reduced. }
	\end{center}
\end{figure} 

We should now verify the behaviour of the cross section, Eq.~\ref{eq:sigma_lo_nlo} with respect to the remaining factorization scale dependence. Therefore, we set the scale at the LO PDF ($\mu_F = \mu_0$) and in the NLO coefficient $C^{\rm NLO}(\mu_0)$, while varying the  factorization scale, $\mu_F$, in the PDF multiplying the NLO contribution. We will investigate the central prediction $\mu_F = \mu_0$ and also a smaller factorization scale $\mu_F = \mu_0/2$. Here we cannot use the larger $\mu_F = 2 \mu_0$, because it would allow the DGLAP evolution to violate the cutoff. This would happen by the emission of partons with enough transverse momentum to produce a photon with some transverse momentum. Therefore, the dilepton will have to carry this momentum and the net effect will be a reduction of the azimuthal angle $\phi$, putting, in the forbidden region, some events previously understood to be in the allowed region of $\phi$.

In Fig.~\ref{fig:sensitivity}, we obtained the scale variation described above for the differential cross section in rapidity for $M = 6$\,GeV and $12$\,GeV, setting the LHC energy to 14 TeV and, as an example, applying the azimuthal angle cutoff $\phi_0 = 0.85\pi$ with $\mu_0 = 0.44 M$. The renormalization scale is kept fixed at $\mu_R = M$. We can see that changing the factorization scale does not change much the results. Therefore, the role of optimal scale still holds and the uncertainty in the choice of scale is reduced. 

%-----------------------------------------------------------------
\section{Conclusion}
\label{Conclusion}

In this work, we investigated the production of Drell--Yan dileptons at small $x$ with a cutoff that excluded smaller values of the azimuthal angle $\phi<\phi_0$. Following the prescription established in earlier works, we calculated the leading order optimal factorization scale using the dominant diagram at NLO, i.e., the gluon--quark Compton scattering. In doing so, the main theoretical uncertainty (factorization scale) was reduced, as it can be seen for $\phi_0 = 0.85 \pi$ at Fig.~\ref{fig:sensitivity}.

We provided the optimal scale as function of the size of the cutoff $\phi_0$ in Fig.~\ref{fig:mu0}. By introducing the cutoff, it was possible to lower the scale at which the parton distributions are probed, for example, $\mu_0=0.44M$ at $\phi_0 = 0.85 \pi$. In order to avoid the DGLAP evolution of the PDFs spoiling the proposed observable by the emission of a parton in the cutoff region, appropriate Sudakov factors were included. They changed the dependence of the optimal scale on $\phi_0$, but for $\phi_0 > 0.7 \pi$, the change of its absolute value is very small and therefore the optimal scale is quite robust regarding this correction. 

Finally, we calculated the cross section of the discussed observable with $\phi_0 = 0.85 \pi$  in Fig.~\ref{fig:uncert}, showing that indeed we will have a smaller cross section by a factor of about 2 when compared with the case without the cutoff. The uncertainty bands shown indicate that the determination of the parton distributions can be improved, since the uncertainty due to the factorization scale was greatly reduced.

%-----------------------------------------------------------------
\section*{Acknowledgements}

We thank very much Alan Martin and Misha Ryskin for fruitful discussions. This work was supported by Fapesc, INCT-FNA (464898/2014-5), and CNPq (Brazil) for MKY and EGdO. This study was financed in part by the Coordena\c{c}\~ao de Aperfei\c{c}oamento de Pessoal de N\'ivel Superior -- Brasil (CAPES) -- Finance Code 001.

%-----------------------------------------------------------------

%-----------------------------------------------------------------
\end{document}